%% file: main.tex
\documentclass[sigconf, 10pt]{acmart}
\AtBeginDocument{%
  }

\setcopyright{none}

\renewcommand\footnotetextcopyrightpermission[1]{}
\usepackage{color,soul}

\begin{document}

\title{Are We There Yet? A Measurement Study of Efficiency for LLM Applications on Mobile Devices}


\author{Xiao Yan}
\affiliation{%
  \institution{The University of Texas at Dallas}
  \city{Richardson}
  \country{USA}}
\email{xiao.yan@utdallas.edu}

\author{Yi Ding}
\affiliation{%
  \institution{The University of Texas at Dallas}
  \city{Richardson}
  \country{USA}}
\email{yi.ding@utdallas.edu}

\renewcommand{\shortauthors}{Xiao Yan and Yi Ding}

\begin{abstract}
Recent advancements in large language models (LLMs) have prompted interest in deploying these models on mobile devices to enable new applications without relying on cloud connectivity. However, the efficiency constraints of deploying LLMs on resource-limited devices present significant challenges. In this paper, we conduct a comprehensive measurement study to evaluate the efficiency tradeoffs between mobile-based, edge-based, and cloud-based deployments for LLM applications. We implement AutoLife-Lite, a simplified LLM-based application that analyzes smartphone sensor data to infer user location and activity contexts. 
Our experiments reveal that:
(1) Only small-size LLMs (<4B parameters) can run successfully on powerful mobile devices, though they exhibit quality limitations compared to larger models;
(2) Model compression is effective in lower the hardware requirement, 
but may lead to significant performance degradation;
(3) The latency to run LLMs on mobile devices with meaningful output is significant (>30 seconds), while cloud services demonstrate better time efficiency (<10 seconds);
(4) Edge deployments offer intermediate tradeoffs between latency and model capabilities, 
with different results on CPU-based and GPU-based settings.
These findings provide valuable insights for system designers on the current limitations and future directions for on-device LLM applications.
\vspace{-8pt}
\end{abstract}



\vspace{-8pt}
\keywords{On-Device Learning, Foundation Model, Real-Time}
\vspace{-8pt}


\maketitle

\input{Sec1-Intro}

\input{Sec-Application}

\input{Sec3-Measurement}

\input{Sec5-Evaluation}

\input{Sec-Related-Work}
\input{Sec6-Discussion}

\bibliographystyle{ACM-Reference-Format}
\bibliography{sample-base}


\end{document}

%% file: Sec1-Intro.tex
\vspace{-5pt}
\section{Introduction}
In recent years, large language models (LLMs) and vision language models (VLMs) have been greatly advanced and applied in various domains.
Given their impressive reasoning and generating capacities, 
some recent work has explored the LLMs and VLMs to perceive the environment through smartphone sensors and conduct inferences, such as SHARE~\cite{zhang2023unleashing}, PenetrativeAI~\cite{xu2024penetrative}, and AutoLife~\cite{xu2024autolife}.
However,
in most existing work, LLMs and VLMs are running remotely on the cloud server,
which incurs some potential limitations: 
dependence on stable network access, 
uncontrolled latencies due to network and server status, 
and privacy risks in uploading data.
Therefore,
deploying the LLMs and VLMs on local mobile devices can enable new applications in broader scenarios by overcome these limitations.

In the mobile computing community,
research work have been done from different perspectives to enable training and prediction on mobile devices (latency~\cite{jiang2021flexible, zhang2021nn, guo2021mistify, yi2023boosting},
memory~\cite{gim2022memory, yang2022rep, kim2023device}, and energy~\cite{liu2018demand, jia2022codl, rastikerdar2024cactus}).
The solutions can be categorized as 
model compression and selection~\cite{gim2022memory, yang2022rep, kim2023device, rastikerdar2024cactus},
federated learning~\cite{wang2023eefl, chen2022fedsea},
and  heterogeneous computing~\cite{jia2022codl}.
However,
the existing work is mostly focused on the traditional deep-learning framework,
which cannot address all the challenges in deploying LLMs and VLMs.
For the topic of LLMs on mobile and edge devices,
survey and position papers published recently laid the foundation of this emerging direction~\cite{chen2025towards, zheng2024review}.
Specific works focus on the different perspectives, 
including privacy and security~\cite{yuan2024wip, li2024governing},
model customization and personalization~\cite{zhuang2024litemoe, qin2024enabling},
and model/system optimization~\cite{yu2024edge, ding2024enhancing, xu2025fast}.
However,
most work does not provide a horizontal comparison of the system efficiency across different mobile, edge, and cloud platforms.

In this paper,
by conducting an experimental study on the efficiency of different LLM deployments (i.e., mobile, edge, and cloud) for a mobile application,
we aim to provide a holistic comparison and discussion to enhance the community's understanding of potential tradeoffs of different deployment settings.
Specifically,
we implemented a simplified version of \texttt{AutoLife}~\cite{xu2024autolife},
a recent work that uses multi-modality sensor data on smartphones to infer human locations and activities and generate a diary for the users.
We deploy the system with three different settings: mobile-based, edge-based, and cloud-based,
and compare the memory consumption and the latency of the system.

The major observations and conclusions we have include:
(1) We successfully deployed four small-size LLMs (Gemma-2B, Gemma2-2B, Llama3.2-1B, and Llama3.2-3B) on an Android phone with GPU and 8GB RAM,
but only one model (Gemma2-2B) provides meaningful answers.
Meanwhile,
all LLMs with 7B+ parameters can provide meaningful answers,
but these models cannot be deployed on mobile devices due to limited memory.
This indicates the design space is limited for systems that require running LLMs locally.
(2) We identified the drawbacks of model compression.
The compressed versions of Llama3.2 deployed on mobile devices fail to generate meaningful answers, 
while the original versions work on edge servers.
(3) The latency to run LLMs on mobile devices with meaningful output is 
significant (i.e., >30 seconds), 
while using a cloud service with API is more time-efficient (i.e., <10 seconds).
For all deployments,
the latency positively correlates with the model size regardless of model series (e.g., DeepSeek or Llama),
but different model series with the same model sizes have different latencies.
(4) A typical one-GPU-based edge server shows much higher efficiency (i.e., higher model output speed, lower latency with smaller variance) than a typical 8-core-CPU-based edge server,
indicating the superiority of GPU-based servers for LLM-based applications.
(5) The model's speed (i.e., number of tokens output per second) negatively correlates with the model size,
indicating the difficulties of adopting large-size LLMs for real-time applications.

The contribution of the work is three-fold:
(1) We deployed an LLM-based mobile application in three different settings (i.e., mobile, edge (CPU-based and GPU-based), and cloud) and measured the latency and memory consumption to provide an aligned comparison.
The code used in the paper will be published so that the researchers can use it to conduct the following work.
(2) We conducted thorough experiments with different model versions (Gemma, Llama, DeepSeek, Qwen, GPT, Claud) and different model sizes (e.g., 0.5B, 1B, 2B, 3B, 4B, 7B, 8B) to evaluate the system efficiency.
(3) We obtained some interesting observations and conclusions from the experiments,
which can help enhance the community's understanding of the practicability and potential challenges of mobile LLM applications.




%% file: Sec-Application.tex
\vspace{-5pt}
\section{LLM Application}


The adoption of LLMs has been an emerging topic in the CPS-IoT community~\cite{baris2025foundation}.
Applications explored include
spatiotemporal data mining~\cite{ouyang2024llmsense, jin2023time},
mobile tasking~\cite{liu2024tasking, wen2024autodroid, yuan2024mobile, lee2023explore, cosentino2024towards},
and mobile sensing and reasoning~\cite{xu2024penetrative, xu2024autolife, cosentino2024towards, an2024iot}.
In this work,
we choose to implement a key component in \texttt{AutoLife}~\cite{xu2024autolife}, a novel application to use smartphone sensors (i.e., GPS, Wi-Fi, IMU, Barometer) to infer user location/context and compile daily diaries.
The motivation to implement \texttt{AutoLife} are two folds:
(1) The application's task is complex enough to unveil the potential of mobile LLM and simple enough to be solved with small LLMs (e.g., Gemma2-2B).
Actually, 
our results indicate that the task in \texttt{AutoLife} is at the boundary of solvable (at least 2B model needed) and executable (at most 3B can be deployed on smartphones with 8 GB memory) using mobile LLM.
(2) Multiple types of sensor data are collected and processed to illustrate LLMs' capacity in sensor data understanding and reasoning,
which can further motivate future work in LLM for perception.

\begin{figure}[ht]
   \vspace*{-3pt}
    \centering
    \includegraphics[width=0.95\linewidth]{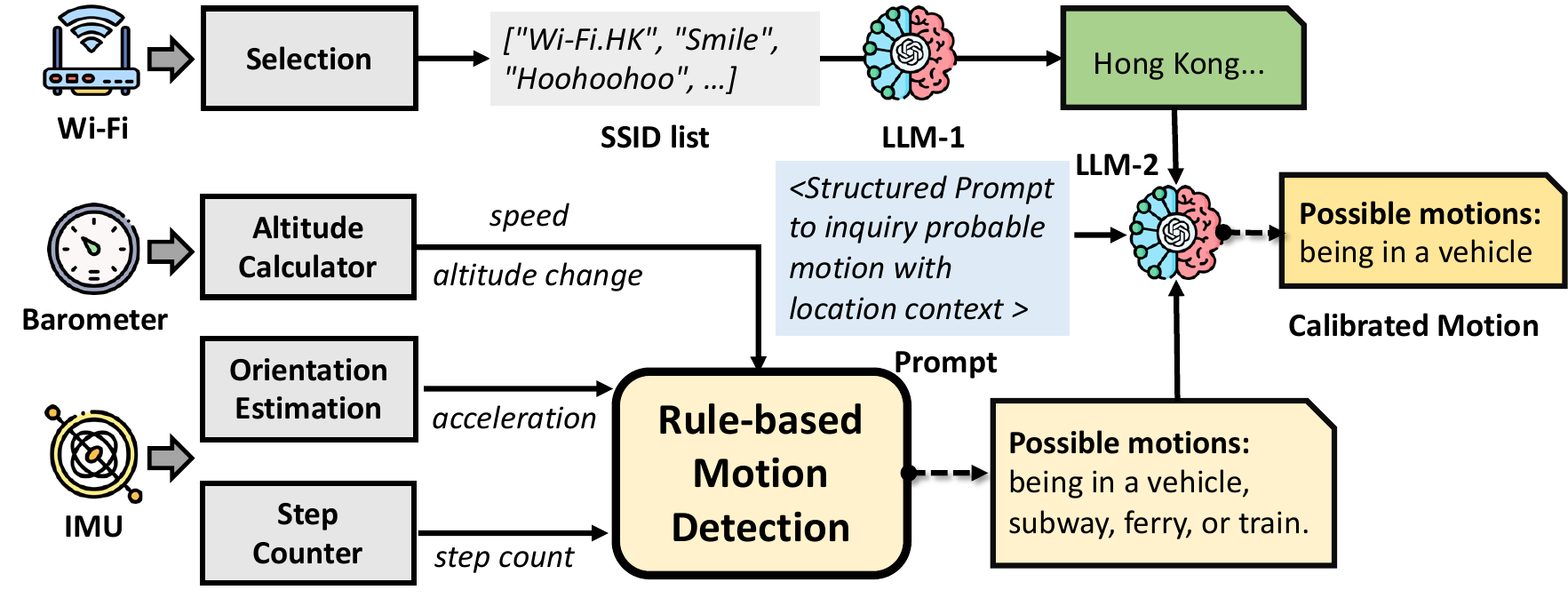}
    \vspace*{-10pt}
    \caption{AutoLife-Lite}
    \label{fig:autolife}
    \vspace*{-9pt}
\end{figure} 

We implement \texttt{AutoLife-Lite} (Figure ~\ref{fig:autolife}),
a lite version of the location and motion context fusion part in \texttt{AutoLife}~\cite{xu2024autolife} without the VLM module because there is no available VLM to deploy on mobile devices.
The subtask implemented in \texttt{AutoLife-Lite} is to use the Wi-Fi, barometer, and IMU data to infer the location and motion of the user,
which is a key module in \texttt{AutoLife}.
Specifically,  
the Rule-based Motion Detection collects sensor data by registering listeners with Android's sensor system. 
When a sensor reports new data, the detector processes it through callback functions that update internal variables. 
This collected data is combined in a detection algorithm that uses predefined thresholds to classify the user's current motion state, such as walking, running, or riding in a vehicle.

%% file: Sec3-Measurement.tex
\begin{figure}[ht]
   \vspace*{-7pt}
    \centering
    \includegraphics[width=0.8\linewidth]{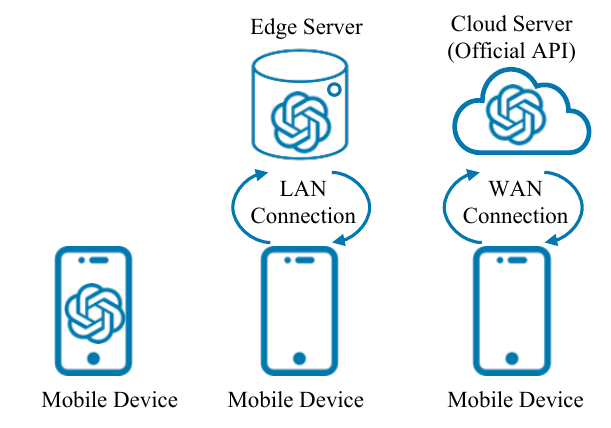}
    \vspace*{-10pt}
    \caption{Mobile-, Edge-, and Cloud-based Deployment}
    \label{fig:deployment-setting}
    \vspace*{-8pt}
\end{figure} 

\vspace{-5pt}
\section{Application Deployment}
We envisioned three typical deployment settings of using LLMs in mobile applications as illustrated in Figure~\ref{fig:deployment-setting}:
\textbf{Mobile-based}, \textbf{Edge-based}, and \textbf{Cloud-based}.
In \textbf{Mobile-based} deployment,
the LLM models (usually a small-sized or compressed model like 2B) are completely on mobile devices
to ensure data privacy and usability without a network connection.
In the \textbf{Edge-based} deployment,
LLM models (a medium-sized model like 8B) are deployed on the edge server,
where mobile devices can connect to the server with a local area network (LAN) (e.g., Wi-Fi, cellular).
A potential scenario is a group of drones as mobile devices connect to the edge server on the cellular station using a cellular network.
In the \textbf{Cloud-based} deployment,
the LLM models are deployed by commercial companies on cloud servers, and APIs are provided to access the service through wide area network (WAN) (e.g., Internet).
Most of the existing mobile APPs are deployed in this fashion because 
(1) availability of latest models;
(2) minimum deployment effort;
(3) minimum memory consumption on mobile devices needed.

In this work,
we deployed \texttt{AutoLife-Lite} on all three settings to provide a holistic observation of the performance and efficiency (i.e., latency, memory) of different deployments.

\vspace{-8pt}
\subsection{Mobile-based Deployment}
\noindent \textbf{Hardware:}
We use a Google Pixel 8~\cite{GooglePixel8Specs} as the mobile device,
which is equipped with 8GB RAM,
a Google Tensor G3 processor~\cite{WikipediaGoogleTensor} 
(with a 10-Core GPU Immortalis-G715s MC1 and Nona-core CPU),
and a customized TPU.

\noindent \textbf{Software and Framework:}
We leverage Google MediaPipe~\cite{GoogleMediaPipeGuide} (recently rebranded as LiteRT) to deploy Gemma-2B and Gemma2-2B 
and PyTorch Executorch \cite{PyTorchExecutorch} to deploy the Llama3.2-1B and Llama3.2-3B on the smartphone.


\vspace{-8pt}
\subsection{Edge-based Deployment}
\noindent \textbf{Hardware:}
We use the Xen-virtualized server~\cite{AWS_EC2_P3_Instances} the AWS platform provides as the edge server.
We use different settings to simulate a CPU-based and a GPU-based server.
For the CPU-based server, we used 800\% CPU resources with 500 GB memory, which represents eight full cores of processing power on an Amazon P3 instance with Intel Xeon E5-2686 v4 processors.
For the GPU-based server, we used one Tesla V100-SXM2-16GB GPU with 16GiB memory.

\noindent \textbf{Software and Framework:}
We use Ollama~\cite{OllamaWebsite}, a lightweight and extensible framework for building and running language models on the local machine (e.g., macOS, Linux, and Windows). 
The LLMs hosted include Llama (3.2-3B, 3.2-1B, 3.1-8B), DeepSeek-R1 (1.5B, 8B), Qwen (0.5B, 1.8B, 4B, 7B), Gemma (2-2B, 2B, 7B).

\vspace{-8pt}
\subsection{Cloud-based Deployment}
We use Wi-Fi to access the Internet, and HTTP requests are used to call the APIs provided by the companies. 
The LLMs used include 
Claude (claude-3-sonnet-20240229-70B, claude-3-haiku-20240307-20B), 
OpenAI GPT (gpt-4-turbo-preview-1700B, gpt-4o-mini-8B), 
and Google Gemini (gemini-1.5-pro-1500B, gemini-2.0-flash-40B).


%% file: Sec5-Evaluation.tex
\vspace{-3pt}
\section{Evaluation}
We evaluate the efficiency performance (i.e., memory consumption, latency) of the three deployments (i.e., mobile-based, edge-based, and cloud-based) of \texttt{AutoLife-Lite}.
We did not measure the accuracy of model output as it's outside the scope of this paper,
but we have conducted manual validation of the output to identify cases in which the model failed to provide the output or provided hallucinated output.

\vspace{-5pt}
\subsection{Measurement Methodology}
\noindent \textbf{Memory Consumption Measurement.}
We measured the smartphone memory consumption in all three deployments to provide an aligned comparison.
We use resident set size (RSS)~\cite{WikipediaRSS} to directly access the process status file (/proc/<pid>/status) and measure the total physical memory held in RAM. 
This metric represents the actual memory footprint of our application in the Android system. 
All the results are based on the setting that \texttt{AutoLife-Lite} is the only APP running and runs in the foreground.
We also measured the memory consumption of the edge server in the edge-based deployment.
For memory on the edge server,
we employ a dual-phase approach to measure memory utilization during LLM inference. For CPU memory (RAM), we capture baseline measurements before inference begins and peak measurements upon completion, with the difference representing the model's RAM footprint. 
Simultaneously, we track GPU memory using NVIDIA's SMI tools, measuring allocated and actively used memory for each GPU. 
A GPU is considered ``active'' when its utilization or memory usage exceeds a certain threshold (5\% in our work based on empirical observation).
Throughout the inference process, 
a dedicated monitoring thread samples all resources at 500ms intervals, 
providing visibility into memory allocation patterns, load distribution across multiple GPUs, and transient resource spikes. 
This comprehensive approach enables detailed analysis of memory requirements across different model architectures and sizes.


\noindent \textbf{Latency Measurement.}
We measure the latency to evaluate if LLM-based applications like \texttt{AutoLife-Lite} can be achieved in real-time and what the impact of different deployment settings and model sizes on the latency.
Specifically,
in the mobile-based deployment,
the latency is measured as the duration between the time that sensor data is collected and the time the local model on the smartphone generates the inference results.
We use System.currentTimeMillis() to capture the timestamps for consistent measurement across all operations.
In the edge-based deployment,
the latency is measured as the duration between the time that the sensor data is received on the edge server and the time the local model on the edge server generates the inference results.
The time is obtained from the log of Ollama.
Note that the data transmission time is not included since 
(1) the transmission time is negligible compared to the model processing time;
(2) the transmission time may vary significantly in different settings.
In the cloud-based deployment,
the latency is measured as the duration between the time that sensor data is collected and the time the inference results are transmitted back to the device. 
Note that the data transmission time is included in the cloud since it's difficult to isolate it.
Moreover,
other unknown processing times (e.g., query queuing time on the cloud server) are also included.


\begin{figure}[ht]
   \vspace*{-0pt}
    \centering
    \includegraphics[width=0.9\linewidth]{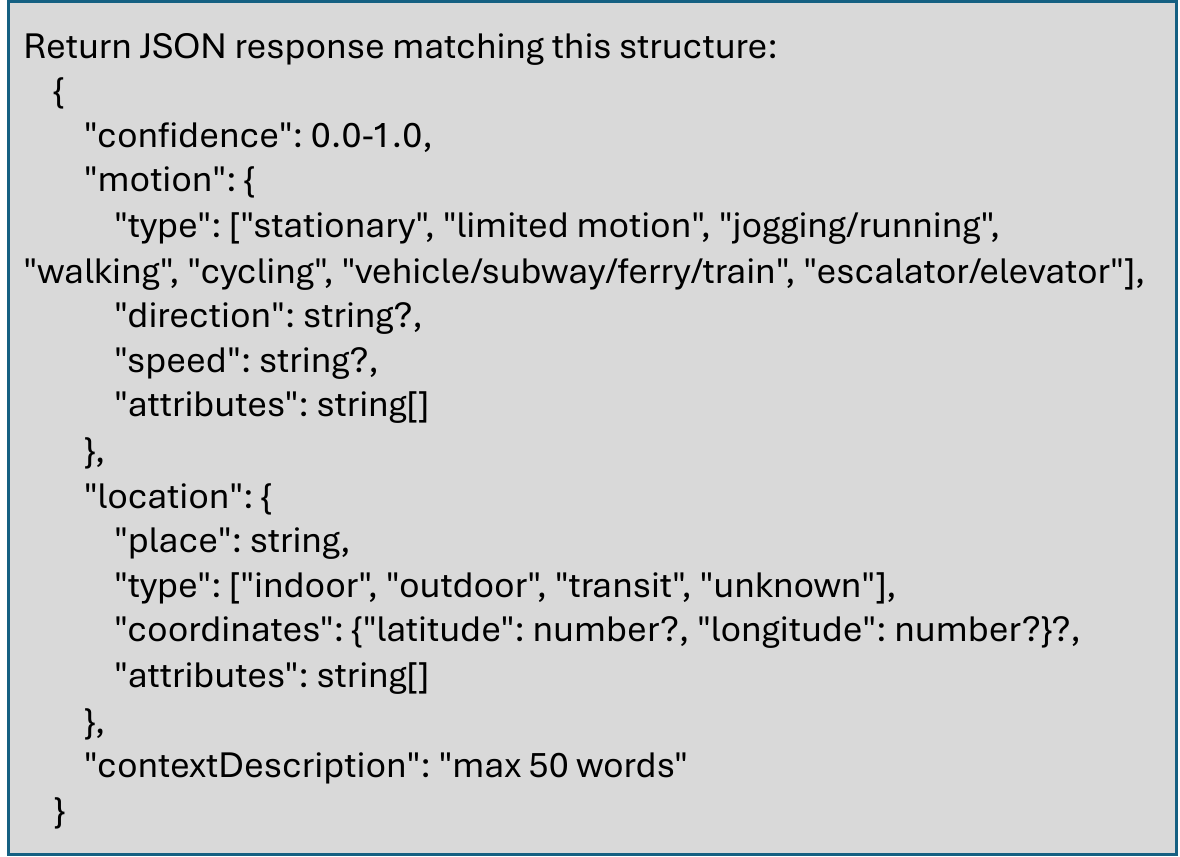}
    \vspace*{-10pt}
    \caption{Structured Prompt to Inquiry Motion}
    \label{fig:prompt}
    \vspace*{-5pt}
\end{figure} 
\noindent \textbf{Structured Prompt.}
To achieve a fair comparison between different models in different deployments,
we use a structured prompt to enforce unified output (Figure~\ref{fig:prompt}).

\begin{figure}[ht]
   \vspace*{-3pt}
    \centering
    \includegraphics[width=0.99\linewidth]{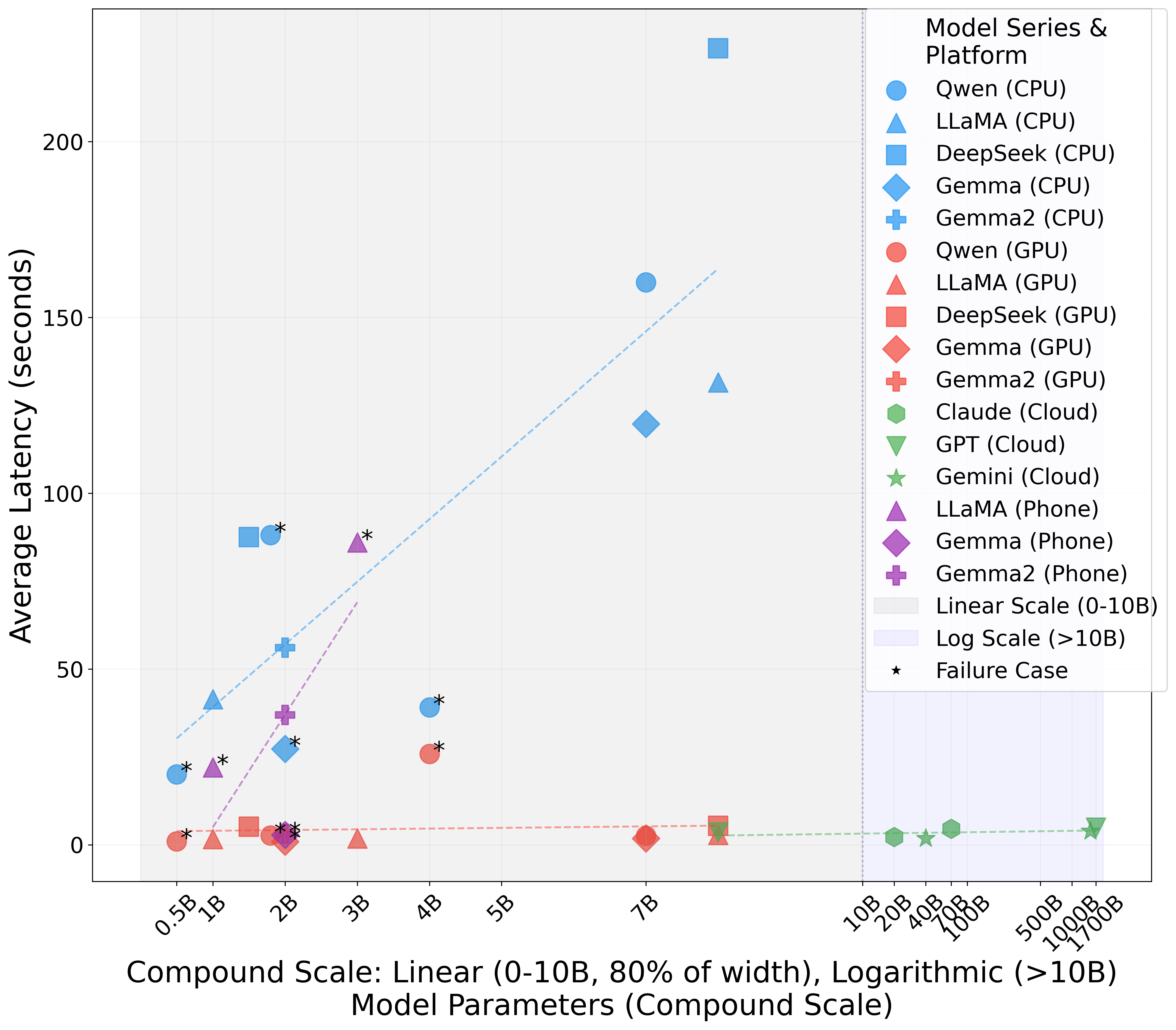}
    \vspace*{-10pt}
    \caption{Model Size v.s. Latency}
    \label{fig:model_size_vs_response_time}
    \vspace*{-7pt}
\end{figure} 

\vspace{-6pt}
\subsection{Evaluation Results\label{subsec:results}}
\noindent \textbf{Overall Results.}
To provide the overall result panoramically,
we use Figure~\ref{fig:model_size_vs_response_time} to illustrate the relation of model size, average latency, and deployment.
We have the following observations:
(1) Cloud-based deployments have the minimum latency (<10s), 
while mobile-based deployments have the maximum latency (>30s for meaningful output),
indicating that cloud-based deployment is still the best way for near-real-time applications. 
(2) GPU-server has smaller and more consistent latency compared to CPU-server,
indicating the importance of GPU in LLM applications.
(3) Only smaller models (<4B) can be deployed on mobile devices, 
but most (3 out of 4) failed to generate meaningful answers.
More detailed results and analysis are as follows.

\noindent \textbf{Model Output Quality.}
Although we didn't conduct thorough experiments to validate the accuracy of the results,
we manually evaluated the results and observed consistent failures for some model-platform combinations.
Specifically,
two types of failures are observed:
(1) Nothing generated: Gemma-2B (Mobile, CPU, GPU).
(2) Failing to generate a reasonable result (i.e., unfinished description or copy-paste of the input): 
Qwen-0.5B (Mobile, CPU, GPU), 
Qwen-1.8B (Mobile, CPU, GPU), 
Qwen 4B (CPU, GPU), 
Llama3.2-1B(Mobile),
Llama3.2-3B(Mobile).
Note that Llama3.2-1B and Llama3.2-3B do not fail on CPU- and GPU-based deployment
because although they are the same model with the same number of parameters,
Llama3.2 on mobile is an optimized version with techniques like LoRA and quantization~\cite{MetaLlama3ModelCard}.
This implies the drawbacks and risks of model compression:
for the same model with the same number of parameters, the non-compressed version works on the edge server, but the compressed version fails on mobile.

\begin{figure}[ht]
   \vspace*{-3pt}
    \centering
    \includegraphics[width=0.99\linewidth]{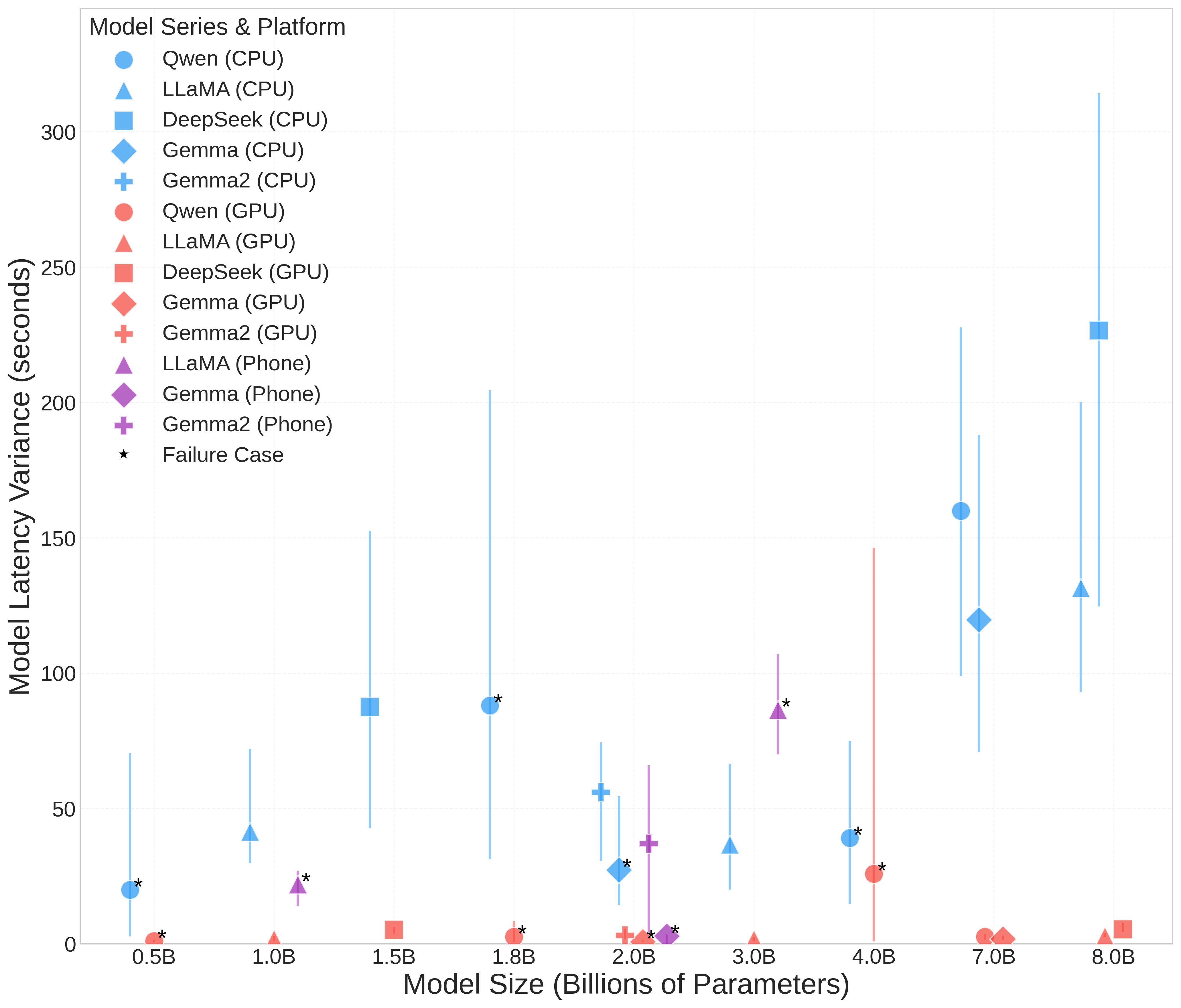}
    \vspace*{-10pt}
    \caption{Model Size v.s. Latency Variance}
    \label{fig:model_size_vs_Latency_Variance}
    \vspace*{-7pt}
\end{figure} 

\noindent \textbf{Latency.}
As shown in Figure~\ref{fig:model_size_vs_Latency_Variance},
the CPU-based edge server has a much higher average latency and larger variance compared with the GPU-based edge server.
A potential reason is that the GPU's specialized memory design makes it efficient for both LLM training and inference.
This result implies the necessity of GPU-based settings for applications with latency requirements.

\begin{figure}[ht]
   \vspace*{-6pt}
    \centering
    \includegraphics[width=0.99\linewidth]{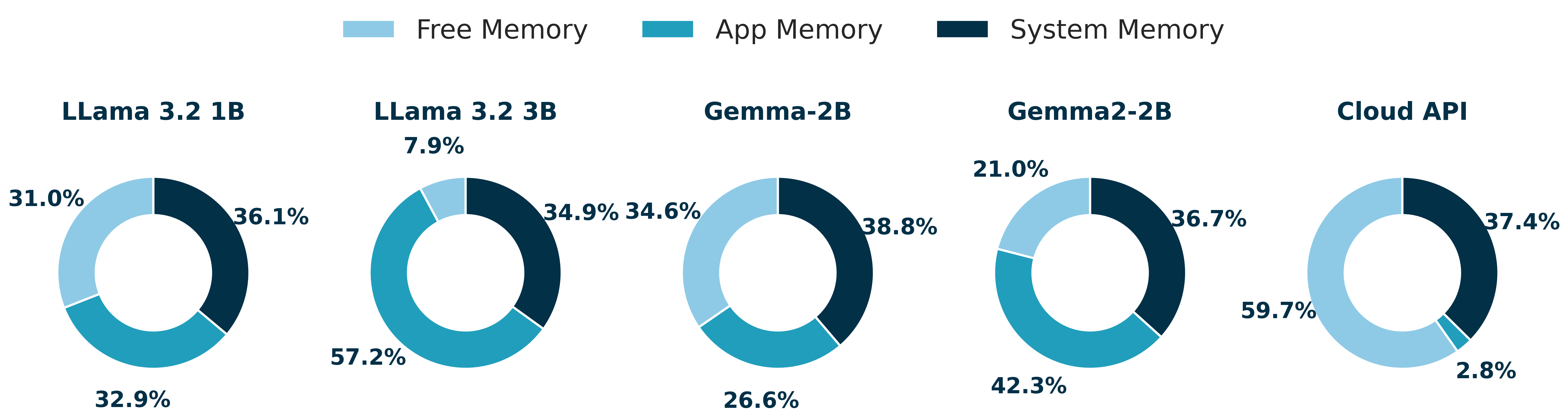}
    \vspace*{-10pt}
    \caption{Memory Consumption on Mobile Device}
    \label{fig:memory-on-mobile}
    \vspace*{-7pt}
\end{figure} 

\noindent \textbf{Memory Consumption on Mobile Devices.}
Figure~\ref{fig:memory-on-mobile} shows a comparison of the memory consumption on mobile devices with mobile-based (left four) and cloud-based (last one) deployment.
Given that the OS consumes around 35\%-40\% memory,
a Llama-3B model would consume 57\% memory,
with only around 8\% free memory.
We use a Google Pixel with 8GB memory in the experiment,
and some recent work has successfully deployed 7B models on smartphones with 16GB memory~\cite{PyTorchXNNPACK}.
As we have verified that models with 7B+  parameters can always produce reasonable output,
smartphone memory becomes the bottleneck for local LLM deployment to wider applications.

\begin{table}[htbp]
\vspace{-5pt}
    \centering
    \small
    \caption{Memory Usage in Edge-based Deployment}
    \vspace{-7pt}
    \begin{tabular}{l@{\hskip 0pt}r@{\hskip 10pt}r@{\hskip 10pt}r}
        \toprule
        \textbf{Model} & \textbf{\shortstack{CPU-based\\(GB)}} & \textbf{\shortstack{GPU-based\\(GB)}} \\
        \midrule
        Qwen-0.5B & 3.04 & 1.88 \\
        Llama3.2-1B & 3.80 & 2.59 \\
        DeepSeekR1-1.5B & 2.50 & 2.00 \\
        Qwen-1.8B & 4.54 & 3.24 \\
        Gemma-2B & 3.38 & 2.82 \\
        Gemma2-2B & 3.88 & 3.41 \\
        Llama3.2-3B & 4.90 & 3.76 \\
        Qwen-4B & 7.08 & 6.02 \\
        Qwen-7B & 10.28 & 9.06 \\
        Gemma-7B & 8.40 & 9.37 \\
        Llama3.1-8B & 6.71 & 6.52 \\
        DeepSeek-R1-8B & 7.01 & 6.52 \\
        \bottomrule
    \end{tabular}
    \label{tab:model_sizes}
    \vspace{-5pt}
\end{table}

\noindent \textbf{Memory Consumption on Edge Servers.}
Memory consumption is similar in CPU- and GPU-based deployment (Table~\ref{tab:model_sizes}).
GPU-based deployment has less memory usage,
as we only measure the usage on GPU memory. Some operations still use CPU even in the GPU-based deployment, and it's hard to count that part.

\begin{figure}[ht]
   \vspace*{-5pt}
    \centering
    \includegraphics[width=0.9\linewidth]{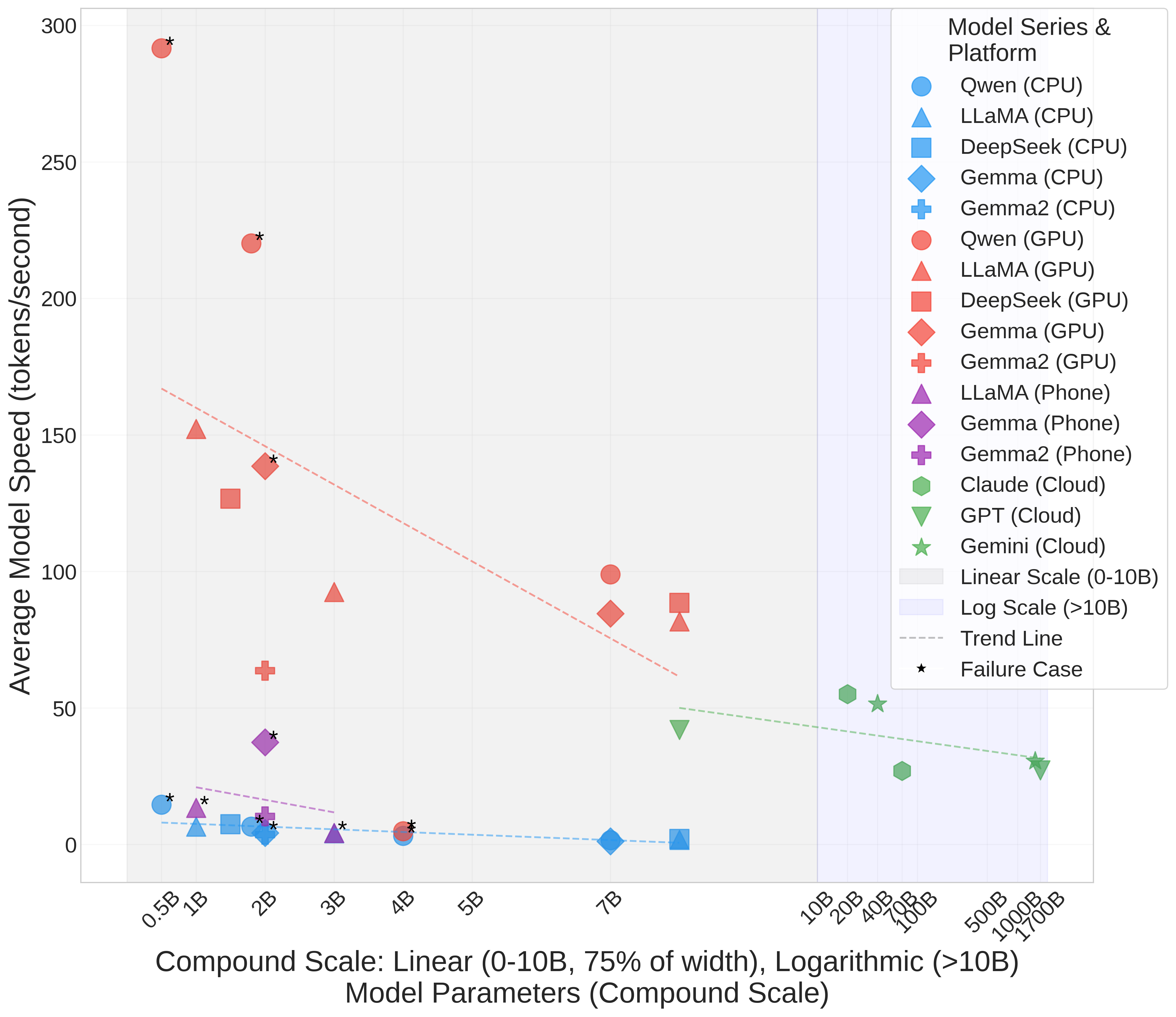}
    \vspace*{-10pt}
    \caption{Model Size and Speed on Edge Server}
    \label{fig:llm-size-vs-speed-on-mac}
    \vspace*{-10pt}
\end{figure}

\noindent \textbf{Model Speed on Edge Server.}
We measure the speed of LLMs running on the edge server and depict its relation with the model size in Figure~\ref{fig:llm-size-vs-speed-on-mac}.
Here, the speed is measured as the ratio of the number of token outputs and time.
The result verifies a straightforward intuition that models with a large number of parameters tend to be slower.
This indicates the challenges of deploying or using large-size LLMs (e.g., >100B) to pursue output quality in a time-efficient way.

%% file: Sec-Related-Work.tex
\vspace{-5pt}
\section{Related Work}

\noindent \textbf{LLM in CPS-IoT.}
The adoption of LLMs has been an emerging topic in the CPS-IoT community~\cite{baris2025foundation}.
Applications explored include
spatiotemporal data mining~\cite{ouyang2024llmsense, jin2023time},
mobile tasking~\cite{liu2024tasking, wen2024autodroid, yuan2024mobile, lee2023explore, cosentino2024towards, lee2024mobilegpt},
and mobile sensing and reasoning~\cite{xu2024penetrative, xu2024autolife, cosentino2024towards, an2024iot}.
The technical challenges solved can be categorized as:
benchmark construction~\cite{quan2024sensorbench, imran2406llasa, yuan2024mobile, an2024iot},
data representation~\cite{wen2024autodroid, xu2024autolife, an2024iot}
models~\cite{yuan2024mobile, an2024iot},
prompting~\cite{ouyang2024llmsense},
memory~\cite{lee2024mobilegpt, wen2024autodroid}.

\noindent \textbf{LLM on Mobile/Edge Devices.}
Several survey and position papers have been published recently to lay the foundation of this emerging direction~\cite{chen2025towards, zheng2024review}.
Some work focuses on the benefit of privacy/security provided by the on-device LLM~\cite{yuan2024wip, li2024governing}.
Some work focuses on the model customization and personalization~\cite{zhuang2024litemoe, qin2024enabling}.
Some work focuses on improving the performance through model or system optmization~\cite{yu2024edge, ding2024enhancing, xu2025fast}
However,
most work does not provide a horizontal comparison of the system efficiency across different mobile, edge, and cloud platforms.
This paper provides a holistic comparison and discussion to help the community understand the potential challenges and tradeoffs when choosing deployment settings for a specific application.

%% file: Sec6-Discussion.tex
\vspace{-5pt}
\section{Discussion}


\noindent \textbf{Limitations and Future Work.}
(1) We only measure the system performance on one mobile device and one edge server,
which may not be typical in other scenarios (e.g., sensor networks),
but the work can still shed light on similar systems with powerful mobile devices and edge servers.
(2) We only measured the efficiency (memory and latency) but did not measure the accuracy.
However, 
we conducted a sanity check of the output and identified some failure cases.
(3) Due to the frameworks and models available, we only test the performance of LLMs but not VLMs. 
On-device VLM is an interesting problem since many mobile devices are equipped with cameras, and videos embed rich information to help understand the environment and human behavior. 
(4) As new technologies developed later may cause our results to appear less accurate,
but we still believe the thorough experimental results reported here can benefit the community in identifying the gaps.

\noindent \textbf{Benchmark and Standards.}
In this work,
we use \texttt{AutoLife-Lite} as a benchmark task to test the LLM efficiency across different deployments.
We are aware of other CPS-IoT benchmarks proposed in the community and plan to study them in the following work.~\cite{quan2024sensorbench, imran2406llasa, yuan2024mobile, an2024iot}.
At the same time,
the heterogeneity in hardware and operating systems brings significant difficulties in comparing different designs.
Platforms like NVIDIA Jetson~\cite{NVIDIAEmbeddedSystems} have the potential to become a standard edge computing platform for AI deployment in resource-constrained environments.



\vspace{-7pt}
\section{Conclusion}
The experimental study reveals significant challenges in deploying LLMs on mobile devices,
including the deployability, output quality, latency, and memory consumption.
These findings suggest that while on-device LLM applications show promise, significant work are needed from different perspectives (e.g., hardware, model, system, application)
before they become practical for real-time applications. 
\vspace{-5pt}